\documentclass[10pt,journal,compsoc]{IEEEtran}

\usepackage{cite}
\usepackage{amsmath,amssymb,amsfonts,upgreek,diagbox}
\usepackage{algorithm}
\usepackage{array}

\usepackage{algorithmic}
\usepackage{cite}
\usepackage{graphicx}
\usepackage{textcomp,makecell}
\usepackage{xcolor}
\usepackage{diagbox}
\usepackage{subfigure}
\usepackage{epstopdf}
\usepackage[justification=centering]{caption}
\usepackage{multirow}
\usepackage{makecell}
\usepackage{enumerate}
\usepackage{url}
\usepackage{color}
\usepackage{diagbox}
\usepackage{CJK}
\usepackage{amsthm}
\usepackage[normalem]{ulem}
\useunder{\uline}{\ul}{}

\newcommand*\ass[1]{[\![#1]\!]}
\newcommand*\mss[1]{\langle#1\rangle}
\newcommand{\RNum}[1]{\uppercase\expandafter{\romannumeral #1\relax}}

\begin{document} 

\title{STR: Secure Computation on Additive Shares Using the Share-Transform-Reveal Strategy}

\author{Zhihua Xia,~\IEEEmembership{Member,~IEEE}, Qi Gu, Wenhao Zhou, Lizhi Xiong, Jian Weng,~\IEEEmembership{Member,~IEEE}, Naixue~N.~Xiong,~\IEEEmembership{Senior Member,~IEEE}

\thanks{Zhihua Xia (corresponding author, email: xia$\_$zhihua@163.com) and Jian Weng are with College of Cyber Security, Jinan University, Guangzhou, 510632, China.} 

\thanks{Qi Gu (email: guqi634337549@163.com), Wenhao Zhou, and Lizhi Xiong are with Engineering Research Center of Digital Forensics, Ministry of Education, School of Computer and Software, Jiangsu Engineering Center of Network Monitoring, Jiangsu Collaborative Innovation Center on Atmospheric Environment and Equipment Technology, Nanjing University of Information Science \& Technology, Nanjing, 210044, China. Lizhi Xiong is also with Guangxi Key Laboratory of Trusted Software, Guilin University of Electronic Technology, Guilin, 541004, China.}

\thanks{Neal N. Xiong is with the College of Intelligence and Computing, Tianjin University, Tianjin 300350, China.}}

\markboth{STR: Secure Computation on Additive Shares Using the Share-Transform-Reveal Strategy}%
{Shell \MakeLowercase{\textit{et al.}}: Bare Demo of IEEEtran.cls for IEEE Journals} 

\IEEEtitleabstractindextext{
\begin{abstract}

The rapid development of cloud computing probably benefits many of us while the privacy risks brought by semi-honest cloud servers have aroused the attention of more and more people and legislatures. In the last two decades, plenty of works seek to outsource various specific tasks to servers while ensuring the security of private data. The tasks to be outsourced are countless; however, the computations involved are similar. In this paper, we construct a series of novel protocols that support the secure computation of various functions on numbers (e.g., the basic elementary functions) and matrices (e.g., the calculation of eigenvectors and eigenvalues) on arbitrary $n\geq 2$ servers. All protocols only require constant rounds of interactions and achieve low computation complexity. Moreover, the proposed $n$-party protocols ensure the security of private data even though $n-1$ servers collude. The convolutional neural network models are utilized as the case studies to verify the protocols. The theoretical analysis and experimental results demonstrate the correctness, efficiency, and security of the proposed protocols.

\end{abstract}

}

\maketitle

\section{Introduction}\label{sec:Introduction}
 
\IEEEPARstart{T}{he} rapid development of digital devices leads us to the era of information explosion, where the local processing equipment becomes unable to meet the needs in many situations. Cloud computing comes to be a perfect solution to this dilemma by providing huge storage and computation resources. For example, the Deep Learning as a Service (DLaaS) could be a promising application paradigm for resource-limited users to utilize AI models that are trained and deployed on the cloud server. At a glance, it seems quite reliable; however, conveniences all come at the cost of exposing privacy as the users generally need to upload their plaintext data to the server for processing \cite{dai2017moved}. In recent years, numerous privacy problems related to cloud servers aroused widespread concern. Accordingly, some government agencies have set up relevant regulations. A well-known example is the General Data Privacy Regulation \cite{voigt2017eu} implemented by the European Union, which clearly defines that the utilization of personal data needs the consent of the data subject. Convenience and privacy are both the public demands. Therefore, how to outsource the computation task without leaking data privacy becomes one of the most urgent problems \cite{gai2016security}.

There are mainly three kinds of schemes that focus on coping with the secure computation outsourcing problem: \emph{differential privacy} \cite{wei2020asgldp}, \emph{Homomorphic Encryption (HE)} \cite{acar2018survey}, and \emph{Secure Multiparty Computation (SMC)} \cite{beaver1991efficient}. Schemes in differential privacy try to ensure security by adding reasonable noise to individual data but retain valid statistical information. However, the significant loss of precision and the task-related perturbation tricks limit their application scenarios \cite{wei2020asgldp}. The HE is another typical cryptography tool that supports the direct computation on ciphertext. Generally speaking, HE schemes are constructed based on computational intractability assumptions \cite{acar2018survey}, which means the encryption and computation process in the schemes always involve operations on large prime numbers, and the raw data can only appear as integers or bits. Therefore, the methods in this category consistently cause unacceptable efficiency and a certain degree of precision loss. Moreover, except the linear computation and multiplication, it is challenging to support other calculations by HE.
 
The above two kinds of schemes can be supported by a single cloud server but have deficiencies in terms of precision or efficiency. After decades of development, more and more cloud computing vendors are willing to provide computation services jointly, which provides a new stage for the SMC schemes. The avoidance of HE makes the recent outsourcing schemes gain thousands of times efficiency advantage \cite{goldreich2019play, huang2019lightweight}. 
 
Additive Secret Sharing (ASS) is one representative SMC technology, which naturally supports the linear operations on ciphertext (i.e., \textit{share}) and has received significant attention. However, most of the existing schemes only support a part of basic elementary functions and are still incompetent when facing the more general operations in practical tasks (e.g., non-linear functions). By introducing Multiplicative Secret Sharing (MSS), Xiong et al. \cite{xiong2020efficient} proposed two resharing protocols to switch the additive and multiplicative \textit{shares}. Then plenty of non-linear protocols are designed with two non-collusion cloud servers based on the resharing protocols. However, the protocols are theoretically designed on an infinite field and need difficult adjustments in practical applications. Besides, the protocols in \cite{xiong2020efficient} can only be conducted on two non-collusion servers; yet, the practical application could be different. For instance, the development of federal learning \cite{kairouz2019advances} has led to the following demand: plenty of different commercial organizations have some valuable data, and they hope to collaboratively train a model (e.g., neural network) without exposing their data. In this case, it is hard to find two servers that are trusted by all participants. Furthermore, the data owners could insist on participating in the secure computation themselves if they do not lack the communication and computing resources. Therefore, it will be difficult to provide secure computation service for real-world challenges with the existing secure computation protocols \cite{xiong2020efficient}.
 
The above demands urge us to improve the protocols further. The protocols should be able to run in a flexibly-defined field with an arbitrary number of servers. Also, the protocols need to be further completed to support more functions. The contributions of this paper can be summarized as follows:
 
\begin{itemize}
 
\item[1)] We propose a ``share-transform-reveal (STR)" strategy to construct secure computation protocols. At first, two novel resharing protocols are designed to transform the additive and multiplicative \textit{shares} to each other with $n$ servers based on STR. Then, the secure computation protocols for all the basic elementary functions are constructed using the resharing protocols. We also design the $n$-party comparison and division protocols with better efficiency by using the STR strategy. All the proposed $n$-party protocols only require constant rounds of interactions.
 
\item[2)] We further design three $n$-party protocols for secure computation on matrix. By processing the matrix as a whole, rather than the separate matrix elements, the interaction rounds of the protocols are independent of matrix dimensions.
 
\item[3)] We prove that the proposed protocols can defend the collusion of $n-1$ servers and present the communication and computation complexities of the protocols in detail. Three convolutional neural network models are further utilized as the case study to demonstrate the efficiency in practice.

\item[4)] The protocols are defined in $\mathbb{R}$, and it is easy for users to map it to a fixed range according to the given task, as shown in the GitHub \cite{code}. Since most of the plaintext tools are defined in $\mathbb{R}$, the further engineer application difficulty can be significantly reduced.

\end{itemize}

The rest of our paper is organized as follows. In Section \ref{sec:related work}, we briefly introduce some typical technologies in secure multiparty computation and secret sharing. The system model, security model, and some preliminaries are given in Section \ref{sec:ProblemFormula} and Section \ref{sec:Preliminaries}. In Section \ref{sec:ReshareProtocol}, we present the STR strategy and two resharing protocols which can transform \textit{shares} between ASS and MSS format. Section \ref{sec:protocols} and \ref{sec:protocolsOnMat} construct protocols for numbers and matrices in our system model. The security and efficiency analysis of all designed protocols are shown in Section \ref{sec:SecuAnaly} and Section \ref{sec:CommComplexity}. Section \ref{sec:CaseStudy} presents a case study of our protocols. Finally, we make the conclusion and prospect in Section \ref{sec:conclu}. Some protocols designed for completeness are placed in the Appendix.

\section{Related work}\label{sec:related work}

\subsection{Secure Multiparty Computation}

Typical SMC schemes enable a group of participants to compute a specific function $\mathcal{F}$$(x_1$, $x_2$, $\cdots$, $x_n)$ collaboratively without revealing their private input $x_i$, where $n$ is the number of participants. SMC is firstly introduced by Yao \cite{yao1982protocols} to cope with the typical Two Millionaires Problem. After decades of development, there are mainly two kinds of schemes that attract plenty of researchers: \emph{Garbled Circuits (GC)} \cite{wu2020efficient} and \emph{Secret Sharing} \cite{beimel2011secret}.

The GC \cite{yao1986generate} introduced by Yao provides a general strategy for $2$-party secure computation. One party generates and encrypts the complete circuit for the functionality $\mathcal{F}$; then the other evaluates it securely with the help of \emph{oblivious transfer} \cite{asharov2013more}. Although GC only needs constant rounds interaction, there are mainly two reasons that GC is not suitable for outsourced computation tasks. On the one hand, the pre-computation is difficult before the task is finalized, and the \emph{offline} (i.e., before the task) consumption is enormous for an actual task \cite{demmler2015automated}. On the other hand, for a GC scheme, each party owns their private data and will get the \emph{knowledge} after the evaluation. However, the servers in outsourced computation protocols are supposed to obtain nothing about private data and \emph{knowledge}. The second reason reflects the difference between the typical SMC scene and the computation outsourcing scene. In this case, the secret sharing technology attracts more researchers in recent years.

\subsection{Secure computation based on Secret Sharing}

 Slightly later than GC, Goldreich \textit{et al.} \cite{goldreich2019play} proposed another typical strategy called \emph{GMW protocol} for secure computation. In \cite{goldreich2019play}, each participant generates a random bit to encrypt (i.e., $XOR$) his input bit (i.e., \textit{secret}). The encrypted bit rather than input attends the computation, and the result is still the encrypted version. The participants can complete the secure computation step by step. Although the interaction rounds are dependent on the depth of task here, the avoidance of complex cryptographic tools (e.g., \emph{oblivious transfer}) and better throughput make the actual efficiency better in the low latency environment \cite{schneider2013gmw}. However, the operations on bits are far from enough to solve the actual task; thus, more recent works \cite{ wagh2020falcon,riazi2018chameleon} focus on the secret sharing schemes that directly support computations on numbers.

 Shamir proposed the first secret sharing scheme in 1979 \cite{shamir1979share}. The author constructed a $(k,n)$-threshold sharing scheme based on \emph{polynomial interpolation}. Here, $(k,n)$-threshold implies that the original \textit{secret} is split into $n$ different \textit{shares}, and with any $k$ \textit{shares} can reconstruct the original \textit{secret} while the any $k-1$ \textit{shares} leak nothing about \textit{secret}. Thus, the $(n,n)$-threshold secret sharing technology is suitable for constructing secure computation outsourcing protocols, where $n \geq 2$. 

 ASS is a typical $(n,n)$-threshold scheme and widely used in secure computation outsourcing. Similar to $XOR$, ASS can support the linear computation without any interactions. This excellent advantage makes researchers consider more complex operations on additive \textit{shares}. Besides the linear computation, multiplication is the other basic operation. Beaver presented a crucial solution called \textit{Beaver triples} in 1991 \cite{beaver1991efficient}, where the additive \textit{shares} of multiplication result can be gotten with only one round of interaction. By extending \textit{Beaver triples}, some later works \cite{ohata2020communication, cryptoeprint:2020:1225} proposed the one-round protocol for secure multiplication with multiple \textit{secrets}.

 Comparison is also a usual operation in computing tasks. Considering the Most Significant Bit (MSB) reflects the sign of \textit{secret}, Damgaard \textit{et al.} \cite{damgaard2006unconditionally} proposed the trick called \emph{bit decomposition} to decompose the arithmetic \textit{share} to its bits secretly. Based on \emph{lagrange interpolation}, the protocol on bits are designed, and the MSB can be further calculated. With a novel correlated randomness trick in the $(2,3)$-threshold, Araki \textit{et al.} \cite{araki2016high} reduced the communication consumption for secure comparison. By introducing a trusted party, Huang \textit{et al.} \cite{huang2019lightweight} further reduced the communication rounds using \emph{carry lookahead adder}. However, the communication complexities (i.e., interaction rounds) of such schemes are kept $O(l)$, where $l$ is the bit-length of \textit{share}. Recently, Wagh \textit{et al.} \cite{wagh2020falcon} proposed a scheme to process all the bits simultaneously, and the result can be revealed by multiplication on each bit. The communication complexity decreased to $O(log_{2}l)$ as the multiplication operations can be conducted in parallel. Obviously, with schemes like \cite{ohata2020communication}, the complexity can be further decreased by the cost of more complex offline work. Yet, the interaction rounds will always be over constant as \cite{ohata2020communication} can not support multiplication on arbitrary numbers of \textit{secrets}.

 The existing works on division protocol are mainly based on two numerical methods: \emph{Goldschmidt's method} \cite{catrina2010secure} and \emph{Newton-Raphson iteration} \cite{ma2019lightweight}. The latter is more recommended due to its higher precision. In these schemes, the dividend range (e.g., integer part of its exponent) should be first calculated. Then with proper initialization, the approximated result can be reached by iteration. With the better comparison protocol, the parallel trick in \cite{wagh2020falcon} accelerated the division process.

 The protocol on non-linear computations (e.g., exponentiation) is also attractive. Based on the inversion trick in \cite{bar1989non}, Damgaard \textit{et al.} \cite{damgaard2006unconditionally} proposed the first protocol on exponentiation. However, the scheme requires high constant interaction rounds. Based on the approximation idea \cite{ma2019lightweight}, tricks like \emph{maclaurin series} are widely used to design non-linear computation protocols. Yu \textit{et al.} \cite{yu2011efficient} noted that the exponentiation computation was essentially the multiplication computation on multiplicative \textit{shares}. Thus, they proposed the resharing protocols to transform the additive and multiplicative \textit{shares}. However, their resharing protocols are based on HE tools, which cause low efficiency and certain precision compromise. Based on a novel allocation of \textit{Beaver triples}, Xiong \textit{et al.} \cite{xiong2020efficient} proposed two novel resharing protocols. The avoidance of HE significantly increases efficiency. Yet, implementation of the protocols needs a very large field in practice. In addition, they still fail to consider some complex protocols like inverse trigonometric functions.

 The abundant protocols attract researchers to apply them in various applications \cite{juvekar2018gazelle, liu2019toward}. However, the schemes in \cite{juvekar2018gazelle, liu2019toward} need to combine the ASS and typical cryptography tools to cope with various real-world challenges. The extra tools cause much higher computation consumption. Some works \cite{huang2019lightweight, wagh2020falcon} are entirely based on ASS and gain much better efficiency. Therefore, the protocol on ASS is a vital bottleneck of such secure computing schemes, and there is still much room for improvement. On the one hand, for the protocols like comparison, the existing schemes still have high communication complexity. On the other hand, approximation-based schemes always cause a certain degree of precision loss. Besides, most previous works pay less attention to a general $(n,n)$-threshold scene $(n>2)$, and the protocols on matrices are rarely considered. Thus, in this work, we focus on the secure computations on the additive \textit{shares} and propose a complete series of protocols for all of the elementary functions and three matrix operations to support various applications with $n\geq2$ servers.

\section{System and security models}\label{sec:ProblemFormula}

 This paper aims to construct secure protocols for computation outsourcing. The problem is how to support accurate computation without privacy leakage. The system and security models are presented as follows.

\subsection{System model}

\begin{figure}[tb]
	\centering
	\includegraphics[width=1.0\linewidth]{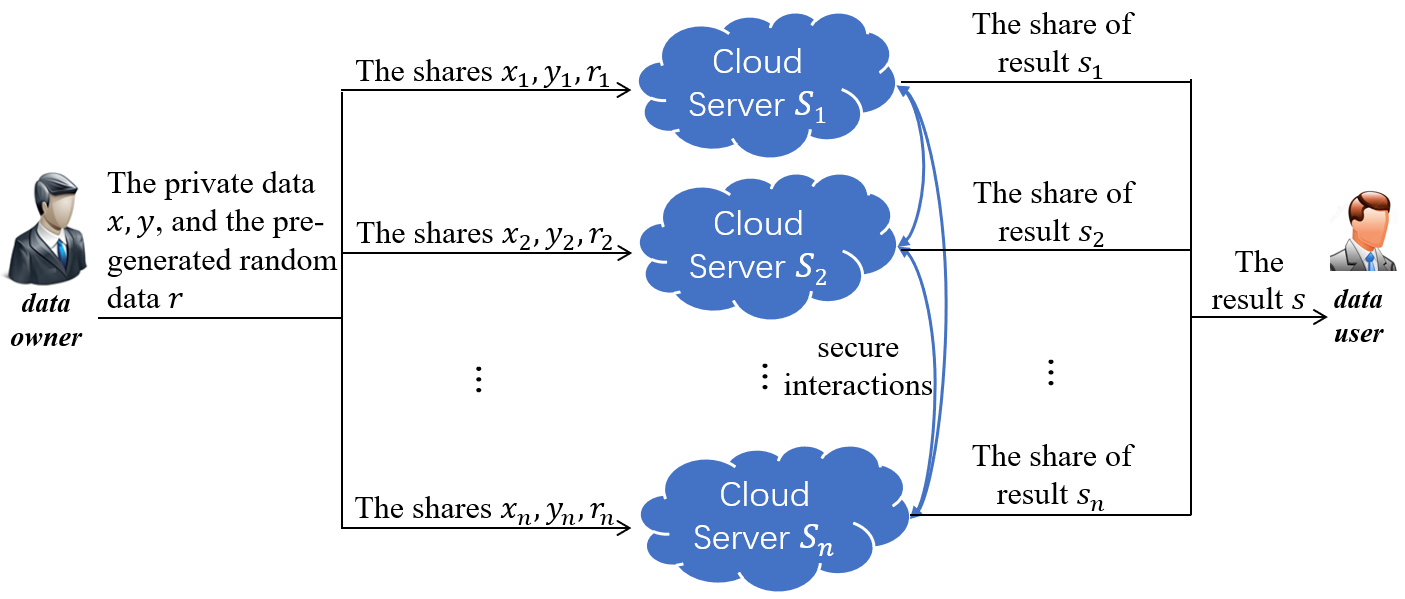}
	\caption{The system model where the $x, y, r$ can be numbers or matrices}
	\label{fig:SystemModel}
\end{figure}

 The proposed system model involves three kinds of entities, i.e., the data owner, cloud servers, and the data user. The system model is shown in Fig. \ref{fig:SystemModel}.

 \textbf{Data owner} owns the original data, i.e., \textit{secret}. We consider the two most common data types: number and matrix in $\mathbb{R}$. At the same time, we assume the data owner is trustworthy to himself, and he will undertake the tasks of generating random numbers and matrices during the \emph{offline phase} (i.e., before the collaborative computation). The \emph{offline phase} task is assumed affordable to the data owner.

 \textbf{Cloud servers} undertake the computation tasks for the data owner during the \emph{online phase} (i.e., the collaborative computation on encrypted data). In this paper, the protocols run on $n$ collaborative servers $S_i$, with $ n \geq 2$ and $i \in [1,n]$. The exact number of servers is unlimited as it is usually determined by the specific task. This paper proposes a series of protocols for the basic computations on numbers and matrices. Then plenty of tasks \cite{ wagh2018securenn, huang2019lightweight,riazi2018chameleon,juvekar2018gazelle} can be executed with better efficiency and precision by combining the protocols.

\textbf{Data user} receives the \textit{shares} of result from each $S_i$ and recovers the real result. In many cases, the data user can also be the data owner.

\subsection{Security model}

 In this paper, the honest-but-curious (also called \textit{semi-honest}) cloud servers are considered. It means each server will execute the protocol as its setting but may attempt to analyze the information from \textit{shares}. As described above, the risk that at most $n-1$ servers collude is considered. Note that the \textit{secret} without encryption will be inevitably exposed when all the servers are involved in the conspiracy; therefore, the considered situation is the worst in practical tasks.

\section{Preliminaries}\label{sec:Preliminaries}

\subsection{Additive secret sharing}

 In ASS, the \emph{secret}, e.g., $x$, is split into $n$ different additive \emph{shares} $\ass{x}_i$, $i \in [1,n]$, satisfying $\sum_{i=1}^{n}\ass{x}_i$ $=$ $x$, where $x \in \mathbb{R}$. In this paper, we focus on the \textit{secret} in the infinite field $\mathbb{R}$ for simplicity and generality. Although a computer (circuit) cannot deal with the infinite field, let us put this problem on hold until subsection \ref{subsec:ActualComputer}.

 ASS is an $(n,n)$-threshold method as the lack of any \textit{share} will make the information independent from \textit{secret}. ASS naturally supports the linear operation: $\sum_{i=1}^{n}(x_i \pm y_i) = x \pm y$. It implies each party can execute $x_i \pm y_i$ locally, and the result is also in the additive form. Similarly, each party can execute multiplication with the constant locally. ASS and other linear secret sharing schemes \cite{beimel1996secure} does not have multiplication homomorphism, which means the interaction is inevitable when executing secure multiplication.

\subsection{Secure multiplication on additive shares}

 To support the multiplication on additive \textit{shares}, Beaver creatively proposed a technique called \textit{Beaver triples} \cite{beaver1991efficient}. The key idea is introducing an \emph{offline phase} to pre-generate a triple $\{a, b, c | c=ab\}$. Then according to the following equation
	\begin{equation}
	\label{eq:SecMul}
	\begin{array}{c}
	(x-a)(y-b) = xy - a(y-b) - b(x-a) - ab,
	\end{array}
	\end{equation}

\noindent where $x, y, a, b \in \mathbb{R}$, the information of $x$ and $y$ will be covered by the triple, but the additive \textit{shares} of $xy$ can be calculated.

\begin{algorithm}[t]
	\caption{Secure multiplication protocol $\texttt{SecMul}$}
	\label{alg: SecMul}
	\begin{algorithmic}[1]
		\REQUIRE
		$S_i$ has $\ass{x}_i, \ass{y}_i$.
		\ENSURE
		$S_i$ gets $\ass{xy}_i$.
		
		\textbf{Offline Phase} :		
		\STATE $\mathcal{T}$ generates random numbers $a$, $b$ $\in$ $\mathbb{R}$, then computes $c$ $=$ $ab$.		
		\STATE $\mathcal{T}$ randomly splits $(a$, $b$, $c)$ into $n$ additive \textit{shares} $(\ass{a}_i$, $\ass{b}_i$, $\ass{c}_i)$ and sends them to $S_i$.		
		
		\textbf{Online Phase} :		
		\STATE $S_i$ computes $\ass{e}_i = \ass{x}_i - \ass{a}_i$ and $\ass{f}_i = \ass{y}_i - \ass{b}_i$.
		\STATE $S_i$ collaboratively recover $e$ and $f$.		
		\STATE $S_i$ computes $\ass{xy}_i = \ass{c}_i$ $+$ $\ass{b}_{i}e$ $+$ $\ass{a}_{i}f$.		
		\STATE $S_1$ computes $\ass{xy}_1 = \ass{xy}_1 + ef$.
		
	\end{algorithmic}
\end{algorithm}

 As shown in Algorithm \ref{alg: SecMul}, a trustworthy party $\mathcal{T}$ is utilized to undertake the task of generating random numbers during the \emph{offline phase}. Each server has one additive \textit{share} of $x$ and $y$. Then, the servers will use $a_i$ and $b_i$ to cover its \textit{shares} and expose $e$ and $f$. According to the Equation (\ref{eq:SecMul}), each server can use $\ass{a}_i$, $\ass{b}_i$ and $\ass{c}_i$ to calculate the additive \textit{share} of $xy$ with $ef$.

\subsection{Multiplicative secret sharing}

 MSS can be seem as a symmetrical technology of ASS. In detail, for a secret $x$, it will be split into $n$ multiplicative \textit{shares} $\mss{x}_i$, $i \in [1,n]$, satisfying $\prod_{i=1}^{n}\mss{x}_i$ $=$ $x$. MSS is an $(n,n)$-threshold in $\mathbb{R}\backslash{}\{0\}$. If the \textit{secret} equals zero, there is at least one party that has the zero \textit{share}, and the \textit{secret} will be exposed to the party. MSS has an important property that $\prod_{i=1}^{n}x_{i}y_{i} = xy$, which means each party can execute multiplication without any interaction. The combination of ASS and MSS would be very usefull in practical applications.

\section{The proposed resharing protocols}\label{sec:ReshareProtocol}

 To utilize both additive and multiplicative homomorphism, we first need to construct the protocols that can transform the \textit{share} between ASS and MSS. Without HE tools, \cite{xiong2020efficient} proposed the first protocols to switch the multiplicative and additive \textit{shares} in the $(2,2)$-threshold. However, the protocols in \cite{xiong2020efficient} can not be adapted for the $(n,n)$-threshold and the direct computation on \textit{shares} leads to an uncontrollable range of value. These two deficiencies impede the application of the protocol in practice.

\subsection{The share-transform-reveal strategy}

 To deal with the two deficiencies in \cite{xiong2020efficient}, we proposed a novel strategy called ``Share-Transform-Reveal (STR)" to construct the resharing protocols. Specifically, the $n$ servers collaboratively \textit{transform} the original \textit{secret} into a new irreversible number with their \textit{shares}. Then, the servers \textit{reveal} the new number collaboratively. Finally, the servers execute the corresponding computation on the new number to generate the wanted \textit{shares}. Both the proposed multiplicative and additive resharing protocols are designed using the STR strategy.

\subsection{Multiplicative resharing}

 When a nonzero \textit{secret} is multiplied by a random nonzero number, it is difficult to infer the \textit{secret} from the result. Following the STR strategy, the proposed multiplicative resharing protocol is described in Algorithm \ref{alg: SecMulReshh}. In \texttt{SecMulResh}, the server $S_i$ holds the multiplicative \textit{share} $\mss{x}_i$ and transforms it to $\mss{\frac{x}{c}}_i$ secretly. Then the servers reveal $\frac{x}{c}$ collaboratively. Finally, the servers can get the additive \textit{share} of $x$ with the help of $\ass{c}_i$. 

\begin{algorithm}[t]
	\caption{Secure multiplicative resharing protocol $\texttt{SecMulResh}$}
	\label{alg: SecMulReshh}
	\begin{algorithmic}[1]
		\REQUIRE
		$S_i$ has $\mss{x}_i$.
		\ENSURE
		$S_i$ gets $\ass{x}_i$.
		
		\textbf{Offline Phase} :
		
		\STATE $\mathcal{T}$ generates random nonzero number $c$ $\in$ $\mathbb{R} \backslash \{0\}$.
		
		\STATE $\mathcal{T}$ randomly splits $c$ into $n$ additive \textit{shares} $\ass{c}_i$ and $n$ multiplicative \textit{shares} $\mss{c}_i$, then sends $(\ass{c}_i$, $\mss{c}_i)$ to $S_i$.
		
		\textbf{Online Phase} :
		
		\STATE $S_i$ computes $\mss{\alpha}_i = \frac{\mss{x}_i}{\mss{c}_i}$.
		
		\STATE $S_i$ collaboratively recover $\alpha$.
		
		\STATE $S_i$ computes $\ass{x}_i = \alpha \times \ass{c}_i$.
		
	\end{algorithmic}
\end{algorithm}

\subsection{Additive resharing}\label{subsec:SecAddReshh}
 
 Symmetric to $\texttt{SecMulResh}$, the additive resharing protocol \texttt{SecAddResh} is shown in Algorithm \ref{alg: SecAddReshh}. The servers firstly transform the $\ass{x}_i$ to $\ass{xc}_i$ by using $\texttt{SecMul}$. Then, we let $S_1$ reveal $xc$, and each server finally gets $\mss{x}_i$ with the help of $\mss{c}_i$.

 Both $\texttt{SecMulResh}$ and $\texttt{SecAddResh}$ are the $(n,n)$-threshold schemes in $\mathbb{R}\backslash\{0\}$. The zero \textit{secret} will be inevitably exposed as at least one multiplicative \textit{share} must equal to zero. We will show that it will not cause actual damage to further protocols in subsection \ref{subsec:discussion}.

\begin{algorithm}[t]
	\caption{Secure additive resharing protocol $\texttt{SecAddResh}$}
	\label{alg: SecAddReshh}
	\begin{algorithmic}[1]
		\REQUIRE
		$S_i$ has $\ass{x}_i$.
		\ENSURE
		$S_i$ gets $\mss{x}_i$.
		
		\textbf{Offline Phase} :
		
		\STATE $\mathcal{T}$ generates random nonzero number $c$ $\in$ $\mathbb{R}\backslash\{0\}$.
		
		\STATE $\mathcal{T}$ randomly splits $c$ into $n$ additive \textit{shares} $\ass{c}_i$ and $n$ multiplicative \textit{share} $\mss{c}_i$, then sends $(\ass{c}_i$, $\mss{c}_i)$ to $S_i$.
		
		\STATE $\mathcal{T}$ generates enough random numbers that the sub-protocol uses and sends them to $S_i$.
		
		\textbf{Online Phase} :
		
		\STATE $S_i$ computes $\ass{xc}_i$ $=$ $\texttt{SecMul}(\ass{x}_i, \ass{c}_i)$.
		
		\STATE $S_i$ sends the \textit{share} $\ass{xc}_i$ to $S_1$, $S_1$ recover $xc$.
		
		\STATE $S_i$ computes $\mss{x}_i = \frac{1}{\mss{c}_i}$.
		
		\STATE $S_1$ computes $\mss{x}_1 = xc\times{}\mss{x}_1$.
		
	\end{algorithmic}
\end{algorithm}

\section{Secure computation Protocols on numbers}\label{sec:protocols}

 In this section, we first design the protocols that support the basic elementary functions based on the above resharing protocols. Then, we further consider the protocols for comparison and division using the STR strategy.

 Elementary functions typically include addition (or subtraction), multiplication, division, exponentiation, logarithm, power, trigonometric functions, inverse trigonometric functions, and their composition. The addition and multiplication can be conducted on additive \textit{shares} as described in Section \ref{sec:Preliminaries}. The protocols for exponentiation, logarithm, power, and trigonometric functions are designed in this section. We discuss the protocol for inverse trigonometric function at Appendix as its construction is more complex. With the proposed protocols, the composition of elementary functions can be executed losslessly in theory.

\subsection{Exponentiation and Logarithm}

 The secure exponentiation protocol tries to securely compute $f(x)=a^x$, where $a$ is the public base number, $a\in(0, 1)\cup(1, +\infty)$. The problem can be converted to calculate  $\ass{a^x}_i$ for $S_i$ with $\ass{x}_i$ as the input, in other words, $\ass{a^x}_i$$\leftarrow$$\ass{x}_i$, $i \in [1,n]$. For simplicity, we omit the definition of the following $i$. According to the equation:

\begin{equation}
\label{eq:SecExp}
\begin{array}{c}
a^{\sum_{i=1}^{n}x_i} = \prod_{i=1}^{n}a^{x_i}, 
\end{array}
\end{equation}

\noindent where $x_i \in \mathbb{R}$, the additive \textit{share} of input can be converted to a multiplicative \textit{share} of the output after each party executes exponential computation. The corresponding protocol is presented in Algorithm \ref{alg: SecExp}.

\begin{algorithm}[t]
	\caption{Secure exponentiation protocol $\texttt{SecExp}$}
	\label{alg: SecExp}
	\begin{algorithmic}[1]
		\REQUIRE
		$S_i$ has $\ass{x}_i$ and public base number $a$. $(a>0$, $a\neq{}1)$
		\ENSURE
		$S_i$ gets $\ass{a^x}_i$.
		
		\textbf{Offline Phase} :
		
		\STATE $\mathcal{T}$ generates enough random numbers that the sub-protocol uses and sends them to $S_i$.
		
		\textbf{Online Phase} :
		
		\STATE $S_i$ computes $\mss{a^x}_i = a^{\ass{x}_i}$.
		
		\STATE $S_i$ collaboratively compute $\ass{a^x}_i = \texttt{SecMulResh}(\mss{a^x}_i)$.
		
	\end{algorithmic}
\end{algorithm}

The situation in logarithm computation, $\ass{log_{a}x}_i$$\leftarrow$$\ass{x}_i$, is symmetric to the exponentiation. According to the equation:

\begin{equation}
\label{eq:SecLog}
\begin{array}{c}
log_{a}\prod_{i=1}^{n}x_i = \sum_{i=1}^{n}log_{a}|x_i|, 
\end{array}
\end{equation}

\noindent where $a \in (0, 1)\cup(1, +\infty)$ and $x \in (0, +\infty)$, the corresponding protocol is presented in Algorithm \ref{alg: SecLog}. In $\texttt{SecLog}$, the additive \textit{shares} are converted to be multiplicative \textit{shares} at first. Then each server computes the logarithm computation with its multiplicative \textit{share}, and the result is rightly the additive \textit{share} of output.

\begin{algorithm}[t]
	\caption{Secure logarithm protocol $\texttt{SecLog}$}
	\label{alg: SecLog}
	\begin{algorithmic}[1]
		\REQUIRE
		$S_i$ has $\ass{x}_i$ and public base number $a$. $(x>0, a>0, a\neq{}1)$
		\ENSURE
		$S_i$ gets $\ass{log_{a}x}_i$.
		
		\textbf{Offline Phase} :
		
		\STATE $\mathcal{T}$ generates enough random numbers that the sub-protocol uses and sends them to $S_i$.
		
		\textbf{Online Phase} :
		
		\STATE $S_i$ collaboratively compute $\mss{x}_i = \texttt{SecAddResh}(\ass{x}_i)$.
		
		\STATE $S_i$ computes $\ass{log_{a}x}_i = log_{a}|\mss{x}_i|$.
		
	\end{algorithmic}
\end{algorithm}

\subsection{Power}

The secure power protocol tries to securely compute $\ass{x^a}_i$$\leftarrow$$\ass{x}_i$, where $a$ is a public number in $\mathbb{Z}$. According to the following equation:

\begin{equation}
	\label{eq:SecPow}
	\begin{array}{c}
	(\prod_{i=1}^{n}x_i)^a = \prod_{i=1}^{n}x_i^a,
	\end{array}
\end{equation}

\noindent where $x_i\in\mathbb{R}$, the power function is entirely based on multiplication homomorphism. Accordingly, the \textit{share} is transformed from ASS to MSS at the beginning and transformed back at last. The protocol is presented in Algorithm \ref{alg: SecPow}. If $a$ is a real number (i.e., non-integer), the computation may becomes a multi-valued function \cite{MultiValuedFunc} and the \emph{complex number} would be involved. For conciseness, this case will be discussed in the Appendix.

\begin{algorithm}[t]
	\caption{Secure power protocol $\texttt{SecPow}$}
	\label{alg: SecPow}
	\begin{algorithmic}[1]
		\REQUIRE
		$S_i$ has $\ass{x}_i$ and public number $a$. ($a$ is integer)
		\ENSURE
		$S_i$ gets $\ass{x^a}_i$.
		
		\textbf{Offline Phase} :
		
		\STATE $\mathcal{T}$ generates enough random numbers that the sub-protocols use and sends them to $S_i$.
		
		\textbf{Online Phase} :
		
		\STATE $S_i$ collaboratively compute $\mss{x}_i = \texttt{SecAddResh}(\ass{x}_i)$.
		
		\STATE $S_i$ computes $\mss{x^a}_{i} = \mss{x}_{i}^a$.
		
		\STATE $S_i$ collaboratively compute $\ass{x^a}_i = \texttt{SecMulResh}(\mss{x^a}_i)$.
		
	\end{algorithmic}
\end{algorithm}

\subsection{Trigonometric functions}

 Trigonometric functions include $sin$, $cos$, $tan$, $cot$, $csc$, and $sec$. The later four functions can be calculated by combining the former two and division operation, and the calculation process of $cos$ is similar to $sin$. Therefore, only the secure computation process of $sin$ is discussed here. For simplicity, we assume that $n$ is odd at first. Then, we have

\begin{equation}
	\label{eq:Sine}
	\begin{array}{l}
	sin(\sum_{i=1}^{n}\theta_i) = \\  
	\sum_{odd \, k \geq 1}(-1)^{\frac{k-1}{2}}\sum_{|A| = k}(\prod_{i\in A}sin\theta_i\prod_{i \notin A}cos\theta_i),
	\end{array}
\end{equation}

\noindent where $\theta_i \in \mathbb{R}$, and $A$ denotes every subset of $[1,n]$. As shown in $\texttt{SecSin}$, since each multiplication in Equation (\ref{eq:Sine}) is combined by the $sin\theta_i$ and $cos\theta_i$, where $i$ denotes every element in $[1,n]$. In this case, each item is a part of the \textit{secret} stored in MSS format. After transforming them into additive \textit{shares}, the $\ass{sin\theta}_i$ has been gotten. When $n$ is even, $sin(\sum_{i=1}^{n}\theta_i)$ can also be calculated by the sum of terms composed by multiplicative \textit{shares}, which means the protocol can be designed similarly.

\begin{algorithm}[t]
	\caption{Secure $sin$ protocol $\texttt{SecSin}$}
	\label{alg: SecSin}
	\begin{algorithmic}[1]
		\REQUIRE
		$S_i$ has $\ass{\theta}_i$.
		\ENSURE
		$S_i$ gets $\ass{sin\theta}_i$.
		
		\textbf{Offline Phase} :
		
		\STATE $\mathcal{T}$ generates enough random numbers that the sub-protocol uses and sends them to $S_i$.
		
		\textbf{Online Phase} :
		
		\STATE $S_i$ computes $sin\ass{\theta}_i$ and $cos\ass{\theta}_i$, please note that one of them is one multiplicative \textit{share} of term (with sign) called as $\{\mss{f^j}_i\}$ in Equation (\ref{eq:Sine}), where $j\in [1, 2^{n-1}]$.
		
		\STATE $S_i$ collaboratively compute $\ass{f^j}_i$ $=$ $\texttt{SecMulResh}(\mss{f^j}_i)$ for all $\mss{f^j}_i$.
		
		\STATE $S_i$ computes $\ass{sin\theta}_i$ $=$ $\sum_{j=1}^{2^{n-1}}\ass{f^j}_i$.
		
	\end{algorithmic}
\end{algorithm}

For inverse trigonometric functions, the \emph{Taylor series} can be used to transform them to be polynomials. It is an acceptable approximation in many situations \cite{xiong2020efficient}. For completeness, we also discuss the corresponding protocol in the Appendix.

\subsection{Secure Computation protocols for Comparison and Division} \label{subsec:CompareAndDivision}

 In Section \ref{sec:ReshareProtocol}, we construct two resharing protocols using STR. The resharing between additive and multiplicative \textit{shares} is useful during handling the non-linear functions (e.g., basic elementary functions) that involve both additive and multiplicative homomorphism. For some multi-variate functions (e.g., comparison and division), it is feasible to compute them similarly by using the resharing protocols \cite{xiong2020efficient}; however, it will bring unnecessary communications. Here we construct the secure computation protocol for comparison and division in a different way by using STR. The particularity of multi-variate functions is that the multiple inputs can cover each other.

 For the comparison, we just need the sign information of the difference between two inputs. As we all know, the multiplication with a positive number will not change the sign of the original number. Here, we let $\mathcal{T}$ provide a random positive number to obscure (\textit{transform}) the difference by $\texttt{SecMul}$. Then, the servers can \textit{reveal} the obscured difference for further computation. The comparison protocol is presented in Algorithm \ref{alg: SecCmp-SCR}.

\begin{algorithm}[t]
	\caption{Secure comparison protocol $\texttt{SecCmp}$}
	\label{alg: SecCmp-SCR}
	\begin{algorithmic}[1]
		\REQUIRE
		$S_i$ has $\ass{x}_i$, $\ass{y}_i$.
		\ENSURE
		$S_i$ gets $sgn(x-y)$.
		
		\textbf{Offline Phase} :
		
		\STATE $\mathcal{T}$ generates a random positive number $t$, then computes the additive \textit{share} $\ass{t}_i$ and sends to corresponding $S_i$.
		
		\STATE $\mathcal{T}$ generates enough random numbers that the sub-protocol uses and sends them to $S_i$.
		
		\textbf{Online Phase} :
		
		\STATE $S_i$ computes $\ass{\alpha}_i = \ass{x}_i - \ass{y}_i$.
		
		\STATE $S_i$ collaboratively compute the $\ass{t\alpha}_i = \texttt{SecMul}($$\ass{t}_i,$ $\ass{\alpha}_i)$.
		
		\STATE $S_i$ collaboratively recover $t\alpha$, the sign of $t\alpha$ is equal to $sgn(x-y)$.
		
	\end{algorithmic}
\end{algorithm}

 Similarly, after multiplied with the same random nonzero number, the ratio between two \textit{secrets} will not change. The dividend can not be equal to zero, and the division protocol is designed as shown in Algorithm \ref{alg: SecDiv-SCR}.

\begin{algorithm}[t]
	\caption{Secure division protocol $\texttt{SecDiv}$}
	\label{alg: SecDiv-SCR}
	\begin{algorithmic}[1]
		\REQUIRE
		$S_i$ has $\ass{x}_i$, $\ass{y}_i$ $(y\neq{}0)$.
		\ENSURE
		$S_i$ gets $\ass{\frac{x}{y}}_i$.
		
		\textbf{Offline Phase} :
		
		\STATE $\mathcal{T}$ generates a random non-zero number $t$, then computes the additive \textit{share} $\ass{t}_i$ and sends them to $S_i$.
		
		\STATE $\mathcal{T}$ generates enough random numbers that the sub-protocol uses and sends them to $S_i$.
		
		\textbf{Online Phase} :
		
		\STATE $S_i$ collaboratively compute the $\ass{tx}_i$ $=$ $\texttt{SecMul}(\ass{t}_i,$ $\ass{x}_i)$, $\ass{ty}_i = \texttt{SecMul}(\ass{t}_i, \ass{y}_i)$.
		
		\STATE $S_i$ collaboratively recover $ty$.
		
		\STATE $S_i$ computes $\ass{\frac{x}{y}}_i = \frac{\ass{tx}_i}{ty}$.
		
	\end{algorithmic}
\end{algorithm}

\section{Secure computation protocols on matrix}\label{sec:protocolsOnMat}

 For simplicity, we only consider the $d\times{}d$ squared matrices (e.g, $Z$, $X$) with the elements in $\mathbb{R}$, where $d$ is a positive integer. In general, the operations on the matrix can be composed of the calculations on its elements. However, the direct implementation may cause $O(d)$ rounds of interaction as most operations can not be executed in parallel. To overcome this problem, we should consider the functions by considering the matrix as a whole. In the following, we first deal with the \textit{dot production} of the matrix and then focus on two fundamental operations on matrices: \emph{the matrix inversion} and \emph{the calculation of eigenvalue and eigenvector}.

\subsection{Dot production}

 The addition and subtraction of the matrix is just the composition of the number. The dot production of matrices can also be composed by multiplication on numbers; however, the separation of matrix elements' internal relations will cause higher consumption. Due to $\sum_{i=1}^{n}X_i\cdot{}Z = X\cdot{Z}$, the dot production has similar property of constant multiplication. In this case, inspired by \textit{Beaver triples}, previous work \cite{wagh2018securenn} constructed the protocol from the perspective of matrices. We further extend it to $(n,n)$-threshold situation as shown in Algorithm \ref{alg: SecMatMul2}.

\begin{algorithm}[t]
	\caption{Secure matrix multiplication protocol $\texttt{SecMatMul}$}
	\label{alg: SecMatMul2}
	\begin{algorithmic}[1]
		\REQUIRE
		$S_i$ has $\ass{X}_i, \ass{Y}_i$.
		\ENSURE
		$S_i$ gets $\ass{XY}_i$.
		
		\textbf{Offline Phase} :
		
		\STATE $\mathcal{T}$ generates enough random vectors to compose $A$, $B$ and computes $C$=$A$$\cdot$$B$.
		
		\STATE $\mathcal{T}$ randomly splits ($A$, $B$, $C$) into two additive \textit{share} ($\ass{A}_i$, $\ass{B}_i$, $\ass{C}_i$) and sends the \textit{share} to $S_i$.
		
		\textbf{Online Phase} :
		
		\STATE $S_i$ computes $\ass{E}_i = \ass{X}_i - \ass{A}_i$ and $\ass{F}_i = \ass{Y}_i - \ass{B}_i$.
		
		\STATE $S_i$ collaboratively recover $E$ and $F$.
		
		\STATE $S_i$ computes $\ass{XY}_i = E \cdot \ass{B}_{i} $ $+$ $\ass{A}_{i} \cdot F + \ass{C}_i$
		
		\STATE $S_1$ computes $\ass{XY}_1 = \ass{XY}_1 + E\cdot{}F$.
		
	\end{algorithmic}
\end{algorithm}

 $\texttt{SecMatMul}$ is similar to $\texttt{SecMul}$ but with the dot production replacing the multiplication. In $\texttt{SecMatMul}$, the data owner needs to know the dimension of \textit{secret} in the \emph{offline phase}, which is sometimes impossible. In this case, the dot production can only be executed in parallel by $\texttt{SecMul}$.

\subsection{Matrix inversion}

 Inspired by \cite{bar1989non}, we construct the secure computation protocol for matrix inversion using the proposed STR strategy. First, the servers \textit{transform} the original matrix (\textit{secret}) to an irreversible matrix collaboratively. Then, the servers compute the inverse of the new matrix and \textit{reveal} it. Finally, the servers transform it back to the actual result. The detailed process is presented in the Algorithm \ref{alg: SecMatInv}.

\begin{algorithm}[t]
	\caption{Secure matrix inversion protocol $\texttt{SecMatInv}$}
	\label{alg: SecMatInv}
	\begin{algorithmic}[1]
		\REQUIRE
		$S_i$ has $\ass{X^{d\times{}d}}_i$ with $rank(X)=d$.
		\ENSURE
		$S_i$ gets $\ass{X^{-1}}_i$.
		
		\textbf{Offline Phase} :		
		\STATE $\mathcal{T}$ generates enough random vectors that the sub-protocol uses and sends them to $S_i$.
				
		\textbf{Online Phase} :		
		\STATE Each $S_i$ generates a random matrix $\ass{Z^{d\times{}d}}_i$ as an additive \textit{share} of the unknown matrix $Z$.		
		\STATE All $S_i$ collaboratively compute the $\ass{ZX}_i$ $=$ $\texttt{SecMatMul}($ $\ass{Z}_i,$ $\ass{X}_i)$.		
		\STATE All $S_i$ collaboratively recover $ZX$ and computes $(ZX)^{-1}$.
		
		\STATE $S_i$ computes $\ass{X^{-1}}_i = (ZX)^{-1} \cdot \ass{Z}_i$.
		
	\end{algorithmic}
\end{algorithm}

 As shown in $\texttt{SecMatInv}$, each server randomly generates an additive \textit{share} of $Z$ to \textit{transform} the secret $X$. Due to $(Z\cdot{}X)^{-1}\cdot{}\sum_{i=1}^{n}Z_i = X^{-1}$, the inversion of $X$ can be substituted by that of $Z\cdot{}X$. Two-time executions of $\texttt{SecMatMul}$ and the random matrix $Z$ ensure the security and efficiency. Note that it is almost impossible that $Z$ is non-full rank (i.e., the probability is zero if the numbers are in $\mathbb{R}$), and it can be detected and remedied by computing the rank of $Z\cdot{}X$.

\subsection{Calculation of eigenvalue and eigenvector}

 The calculation of eigenvalue and eigenvector in the plaintext domain needs multiple iterations (e.g., QR decomposition). Similarly, the calculation in encrypted domain will cause high interaction rounds ($O(d)$) if it is directly conducted on the matrix elements. This paper designs the secure protocol for the calculation of eigenvalue and eigenvector by using the STR strategy. The protocol is based on the following two Lemmas \cite{SimilarMatrix}.

\vspace{1ex}
 \noindent \textbf{Lemma 1.} \emph{If $A$ and $B$ are similar matrices (i.e., $A\sim B$), and $B=P^{-1}AP$, then $A$ and $B$ have the same eigenvalues. If there is an eigenvector $\vec{x}$ under eigenvalue $\lambda$ of matrix $A$, then $P^{- 1}\vec{x}$ is an eigenvector of $B$ under eigenvalue $\lambda$.}
 
\vspace{1ex}
 \noindent \textbf{Lemma 2.} \emph{If $\lambda$ is an eigenvalue of matrix $A$, then $k\times \lambda$ is an eigenvalue of matrix $k\times A$. If there is an eigenvector $\vec{x}$ under eigenvalue $\lambda$ of matrix $A$, then the eigenvector $\vec{x}$ is also under eigenvalue $k\times{}\lambda$ of matrix $k\times{}A$.}
\vspace{1ex}

 The secure computation protocol for the calculation of eigenvalues and eigenvectors is presented in Algorithm \ref{alg: SecEigenVectors}. In $\texttt{SecMatEigen}$, a random matrix $P$ and a random number $t$ are utilized to \textit{transform} the \textit{secret} $X$. The matrix $Y$, which has corresponding eigenvalues and eigenvectors, will be computed and \textit{revealed}. With the help of $t$ and $Z$, the \textit{share} of eigenvalues and eigenvectors of $X$ can be secretly transformed back. Since the the calculation of eigenvalues and eigenvectors is always more time-consuming than one round of extra communication, only $S_1$ undertakes the actual computation task here.

\begin{algorithm}[t]
	\caption{Secure eigenvalues and eigenvectors calculation $\texttt{SecMatEigen}$}
	\label{alg: SecEigenVectors}
	\begin{algorithmic}[1]
		\REQUIRE
		$S_i$ has $\ass{X^{d\times{}d}}_i$ with $rank(X)=d$.
		\ENSURE
		$S_i$ gets one \textit{share} of eigenvalues $\{\ass{\lambda^j}_i\}$ and one \textit{share} of eigenvectors $\ass{V^{d\times{}d}}_i$, $j\in[1,d]$.
		
		\textbf{Offline Phase} :
		
		\STATE $\mathcal{T}$ generates a random non-zero number $t$, then computes the additive \textit{shares} $\ass{t}_i$ for $S_i$.
		
		\STATE $\mathcal{T}$ generates enough random numbers and vectors that the sub-protocols use and sends them to the servers.
		
		\textbf{Online Phase} :
		
		\STATE $S_i$ generates a random square matrix $\ass{P^{d\times{}d}}_i$.
		
		\STATE $S_i$ collaboratively compute $\ass{P^{-1}}_i$ $=$ $\texttt{SecMatInv}($ $\ass{P}_i)$.
		
		\STATE $S_i$ collaboratively compute $\ass{\frac{1}{t}}_i$ $=$ $\texttt{SecDiv}(1,\ass{t}_i)$.
		
		\STATE $S_i$ collaboratively compute $\ass{Y}_i$ $=$ $\texttt{SecMul}(\ass{t}_i,$ $\texttt{SecMatMul}(\ass{P^{-1}}_i,$ $\texttt{SecMatMul}($$\ass{X}_i,$ $\ass{P}_i)))$.
		
		\STATE $S_i$ sends $\ass{Y}_i$ to $S_1$. Then $S_1$ computes the eigenvalues $\{t\lambda^j\}$ and eigenvectors $F$ of $Y$ and sends back to all $S_i$.
		
		\STATE $S_i$ computes $\ass{\lambda^j}_i$ $=$ $\ass{\frac{1}{t}}_i$ $\times$ $t\lambda^j$ for all $j\in[1,d]$.
		
		\STATE $S_i$ computes $\ass{V}_i$ $=$ $\ass{P}_i \cdot F$.
		
	\end{algorithmic}
\end{algorithm} 

\section{Security analysis}\label{sec:SecuAnaly}

 The security of our protocols is discussed under the typical secure multi-party computation framework \cite{canetti2001universally}. The execution of the protocols mainly involves the interaction between servers, and the interaction process is defined as the real experiment. In the proposed protocols, the honest-but-curious cloud server is the potential adversary $\mathcal{A}$. Here, for simplicity, we directly consider the extreme situation that $n-1$ servers are involved in the conspiracy. In the ideal experiment, a simulator $\mathcal{S}$ is defined as the one that can simulate the view of adversary $\mathcal{A}$ according to the functionality $\mathcal{F}$. The protocols are proved secure once the two experiments are indistinguishable. In short, the security can be proved if the view of corrupted parties (i.e., arbitrary $n-1$ servers) is simulatable \cite{canetti2001universally, bogdanov2008sharemind, mohassel2017secureml}. Without the loss of generality, we assume that $S_n$ is not involved in collusion.
 
 Similar to previous works \cite{mohassel2017secureml}, we summarize the functionality $\mathcal{F}$ as follows. In the honest-but-curious model, a trustworthy \emph{functionality} machine owns the true input information of the protocol and generates random numbers or matrices locally; then, it completes the calculation according to the corresponding protocol. In the ideal experiment, the $view$ of $\mathcal{S}$ will be filled with the calculation results. In the following, we will prove that the $view$ is indistinguishable from that in the real world, and the $view$ will not expose the input information. To simplify the proofs, the following Lemma will be used.
 
\vspace{1ex}
\noindent  \textbf{Lemma 3.} \emph{A protocol is perfectly simulatable if all the sub-protocols are perfectly simulatable \cite{bogdanov2008sharemind}.}
\vspace{1ex}

According to Lemma 3, we only need to prove the security of the proposed protocols separately. The security of the composite protocol is guaranteed by the security of components.

\subsection{The properties of ASS and MSS}

\vspace{1ex}
\noindent \textbf{Theorem 1.} \emph{The element $x+r$ is uniformly distributed and independent from $x$ for any element $x\in\mathbb{R}$ if the element $r\in\mathbb{R}$ is also uniformly distributed and independent from $x$.}
 
\noindent \emph{Proof.} If $r$ $\in$ $\mathbb{R}$ is uniformly distributed and independent from $x$, then so is $x+r$, since $f_r(x) = r + x$ is a bijective mapping for $\mathbb{R}$. $\hfill\qedsymbol$
\vspace{1ex}

\noindent \textbf{Theorem 2.} \emph{The nonzero element $xr$ is uniformly distributed and independent from $x$ for any element $x\in\mathbb{R}\backslash{}\{0\}$ if the element $r\in\mathbb{R}\backslash{}\{0\}$ is also uniformly distributed and independent from $x$.}

\noindent \emph{Proof.} If $r$ $\in$ $\mathbb{R}\backslash{}\{0\}$ is uniformly distributed and independent from $x$, then so is $xr$, since $f_r(x) = r \times x$ is a bijective mapping for $\mathbb{R}\backslash{}\{0\}$. $\hfill\qedsymbol$

The above two theorems prove that ASS and MSS are $(n,n)$-threshold in $\mathbb{R}$ and $\mathbb{R}\backslash\{0\}$.
\vspace{1ex}

\subsection{The security of reshairng protocols}

Here, we discuss the security of $\texttt{SecMul}$, $\texttt{SecMulResh}$, and $\texttt{SecAddResh}$. The analyses of many other protocols are similar to them.
\vspace{1ex}
 
\noindent \textbf{Theorem 3.} \emph{The protocol $\texttt{SecMul}$ is secure in the honest-but-curious model.}

\noindent \emph{Proof.} The view of adversary during executing \texttt{SecMul} is $view = (\{\ass{x}_j\}$, $\{\ass{y}_j\}$, $\{\ass{a}_j\}$, $\{\ass{b}_j\}$, $\{\ass{e}_i\}$, $\{\ass{f}_i\})$, where $i\in[1,n]$, $j\in[1,n-1]$. As described in Theorem 1, the recovery of $x$ requires $\ass{x}_n$. Here $\{\ass{a}_j\}$, $\{\ass{b}_j\}$ are uniformly random which is simulatable. Due to a lack of $\ass{a}_n$, the $\ass{e}_n$ is also totally random and simulatable. The situation of $y$ is similar. Besides, as the output is computed by these random numbers, the $output$ is also uniformly random. Therefore, both $view$ and $output$ are simulatable by the simulator $\mathcal{S}$ and the views of $\mathcal{S}$ and $\mathcal{A}$ will be computationally indistinguishable. Notably, the conspiracy of any $n-1$ servers will result in the same situation. $\hfill\qedsymbol$

\vspace{1ex}
The proof for \texttt{SecMatMul} is similar, and we omit it for simplicity. Since both ASS and MSS are $(n,n)$-threshold in $\mathbb{R}\backslash\{0\}$, the view brought by the $n-1$ collusive servers is random and simulatable. Therefore, in the following proofs, we focus on the view gotten from the communication with a non-conspiracy party (i.e., $S_n$).

\vspace{1ex}
\noindent \textbf{Theorem 4.} \emph{The protocol $\texttt{SecMulResh}$ with nonzero input is secure in the honest-but-curious model.}

\noindent \emph{Proof.} The view brought by communication is $view = (\alpha)$. According to Theorem 2, $\alpha$ is random due to the lack of $\mss{c}_n$. Consequently, the $view$ is simulatable and computationally indistinguishable. $\hfill\qedsymbol$

\vspace{1ex}
Please note that if the \textit{secret} is equal to zero, the adversary can infer it from $\mss{x}_1$. The security of \texttt{SecCmp}, \texttt{SecDiv}, \texttt{SecMatInv}, and $\texttt{SecMatEigen}$ can also be proven similarly. Therefore, we leave their proofs out for simplicity.
 
\vspace{1ex}
\noindent \textbf{Theorem 5.} \emph{The protocol $\texttt{SecAddResh}$ with nonzero input is secure in the honest-but-curious model.}

\noindent \emph{Proof.} Besides those brought by \texttt{SecMul}, the view brought by communication is $view=(xc)$. Due to the lack of $c$, $xc$ is random by Theorem 2. Therefore, the $view$ is simulatable and computationally indistinguishable. $\hfill\qedsymbol$
 
\vspace{1ex}

Notably, when the \textit{secret} is equal to zero, the adversary can infer it from $xc$.

\subsection{The security of elementary function protocols}\label{subsec:discussion}

 As described above, the resharing protocols will leak the zero \emph{secret}. Thus, we further discuss the potential risk of the basic elementary functions in this subsection.

 For exponential computation, the public base number $a$ is assumed in the range $(0,1)\cup{}(1,+\infty)$ in the plaintext domain. Therefore, $a^x$ is always greater than zero, which means that the $\texttt{SecMulResh}$ has no zero input. Hence, $\texttt{SecExp}$ is secure in $\mathbb{R}$. The situation of $\texttt{SecLog}$ is similar. For \texttt{SecSin}, each term in Equation (\ref{eq:Sine}) is just a part of the \textit{secret}. It means even some parts are zero, the \textit{secret} is still secure. Due to the property of Equation (\ref{eq:Sine}), it is impossible that all parts are zero from the perspective of $n-1$ servers. Since $\texttt{SecDiv}$ is secure in $\mathbb{R}$, we assert the trigonometric functions are secure in $\mathbb{R}$. 

 For the power function, one server (e.g., $S_1$) will inevitably know that the real \textit{secret} when the \textit{secret} equals to zero. However, we believe the extreme situation is low-destructive for two reasons. On the one hand, zero \textit{secret} generally exposes no important information. For instance, the $0$ in the neural network is frequently produced from the ReLU function. On the other hand, for the multi-variate functions, the exposure of one input will not infect the others. Whatever, we believe that this extreme situation is easy to handle in specific tasks.

\section{Efficiency analysis}\label{sec:CommComplexity}

\subsection{The implementation in finite field} \label{subsec:ActualComputer}
 In the above sections, we discussed the protocols on the infinite field $\mathbb{R}$; however, the real-world computer can only process the data in a finite field, which inevitably brings two limitations: 1) the computer can only cope with the decimal with limited precision, and 2) the computer can only cope with the number within a certain range. Below we analyze how to compute the \textit{share} under the two limitations.

 For the additive \textit{shares}, the fixed-point representation is suitable as the addition makes no precision loss in this situation. But for the multiplicative \textit{shares}, the multiplication and division are more likely to cause precision loss and overflow. The frequent utility of these operations in protocols could make the whole system vulnerable. According to Equation (\ref{eq:SecExp}), the multiplication and division can be converted to be the addition and subtraction of the exponent. Thus, after choosing a default base (e.g., 2), the multiplication can be perfectly expressed by the exponent of addition. It helps to guarantee both efficiency and precision in real computers.
 
 Here we suggest implementing the protocols with a finite field of fixed-point representation. The integer and decimal parts of the number are discussed separately. Let us consider four parameters, i.e., the integer bit-length of \textit{secret}, the integer bit-length of additive \textit{share}, the decimal bit-length of \textit{secret}, and the decimal bit-length of multiplicative \textit{share}. The first parameter depends on the range of numbers in a given task. For instance, if the user can ensure the numbers are in $(-2^{16}, 2^{16}-1)$, he can set the first parameter to be 17. The second parameter determines the degree of randomness, and it is suggested to be at least two times than the first parameter. The last two parameters determine the precision of calculation and can be suggested to be the same as the second parameter. The last three parameters determine the size of $\mathbb{F}$. More details and the performance evaluations are available on the GitHub \cite{code}.

\subsection{Communication and computation complexities}

 Communication consumption is the actual bottleneck in secret sharing based schemes. In the proposed protocols, the servers mainly need to exchange additive and multiplicative \textit{shares}. The communication consumption is summarized in Table \ref{tab:CommEfficiency}. The amounts of additive and multiplicative \textit{shares} are separately listed as their sizes can be different. The consumption is the sum from all of the $n$ servers.

\begin{table*}[th]
	\caption{Main communication consumption of protocols}
	\centering
	\label{tab:CommEfficiency}
	\begin{tabular}{|c|c|c|c|c|c|c|c|}
		\hline
		\multicolumn{1}{|c|}{\multirow{2}{*}{Protocols}} &
		\multicolumn{2}{c|}{\emph{offine phase}} &
		\multicolumn{3}{c|}{\emph{online phase}} \\ \cline{2-6} 
		\multicolumn{1}{|c|}{} &
		additive \textit{share} &
		multiplicative \textit{share} &
		rounds &
		additive \textit{share} &
		multiplicative \textit{share} \\ \hline
		\texttt{SecMul}    & $3n$                     & $0$  & $1$ & $2n^2-2n$   & $0$ 	   \\ \hline
		\texttt{SecMulResh} & $n$                      & $n$  & $1$ & $0$         & $n^2-n$ \\ \hline
		\texttt{SecAddResh} & $4n$                     & $n$  & $2$ & $2n^2-n-1$  & $0$     \\ \hline
		\texttt{SecExp}    & $n$                      & $n$  & $1$ & $0$         & $n^2-n$ \\ \hline
		\texttt{SecLog}    & $4n$                     & $n$  & $2$ & $2n^2-n-1$  & $0$     \\ \hline
		\texttt{SecPow}    & $5n$                     & $2n$ & $3$ & $2n^2-n-1$  & $n^2-n$ \\ \hline
		\texttt{SecSin}    & $n2^{n-1}$ 	& $n2^{n-1}$ 	 & $1$ & $0$ & $(n^2-n)2^{n-1}$\\ \hline
		\texttt{SecCmp}    & $4n$                     & $0$  & $2$ & $3n^2-3n$   & $0$    \\ \hline
		\texttt{SecDiv}    & $7n$                     & $0$  & $2$ & $5n^2-5n$   & $0$    \\ \hline
		\texttt{SecMatMul} & $3d^2n$ 				  & $0$  & $1$ & $2d^2n-2dn$ & $0$    \\ \hline
		\texttt{SecMatInv} & $3d^2n$ 				  & $0$  & $2$ & $3d^2n-3dn$ & $0$    \\ \hline
		\texttt{SecMatEigen} & $(12d^2+4)n$ & $0$ & $5$ & $3d^2n^2+(6d^2-7d-3)n+3n^2-3n$ & $0$ \\ \hline
	\end{tabular}
\end{table*} 

 Using the above implementation scheme, an overall computation consumption of proposed protocols is shown in Table \ref{tab:CompComplexity}. It is easy to note that only trigonometric function has high computational complexity, and approximation tricks like \emph{Taylor series} can further alleviate the problem in practical tasks.

\begin{table*}[]
	\centering
	\caption{The computation complexities of proposed protocols.}
	\label{tab:CompComplexity}
	\begin{tabular}{|l|c|c|c|c|c|c|}
		\hline
		\multirow{3}{*}{Protocols} & \multicolumn{2}{c|}{\emph{offline phase}}    & \multicolumn{4}{c|}{\emph{online phase}}                                               \\ \cline{2-7} 
		& additive \textit{share} & multiplicative \textit{share} & \multicolumn{2}{c|}{additive \textit{share}} & \multicolumn{2}{c|}{multiplicative \textit{share}} \\ \cline{2-7} 
		& add/sub   & add/sub   & add/sub             & mul       & add/sub   & mul    \\ \hline
		\texttt{SecMul}    & $O(n)$    & 0         & $O(n^2)$            & $O(n)$    & 0         & 0      \\ \hline
		\texttt{SecMulResh} & $O(n)$    & $O(n)$    & 0                   & $O(n)$    & $O(n)$    & 0      \\ \hline
		\texttt{SecAddResh} & $O(n)$    & $O(n)$    & $O(n^2)$            & $O(n)$    & $O(1)$    & 0      \\ \hline
		\texttt{SecExp}    & $O(n)$    & $O(n)$    & 0                   & $O(n)$    & $O(n)$    & 0      \\ \hline
		\texttt{SecLog}    & $O(n)$    & $O(n)$    & $O(n^2)$            & $O(n)$    & $O(1)$    & 0      \\ \hline
		\texttt{SecPow}    & $O(n)$    & $O(n)$    & $O(n^2)$            & $O(n)$    & $O(n)$    & $O(n)$ \\ \hline
		\texttt{SecSin}    & $O(n2^n)$ & $O(n2^n)$ & $O(2^n)$            & $O(n2^n)$ & $O(n2^n)$ & 0      \\ \hline
		\texttt{SecCmp}    & $O(n)$    & 0         & $O(n^2)$            & $O(n)$    & 0         & 0      \\ \hline
		\texttt{SecDiv}    & $O(n)$    & 0         & $O(n^2)$            & $O(n^2)$  & 0         & 0      \\ \hline
		\texttt{SecMatMul} & $O(d^2n)$ & 0         & $O(d^2n^2)+O(d^3n)$ & $O(d^3n)$ & 0         & 0      \\ \hline
		\texttt{SecMatInv} & $O(d^2n)$ & 0         & $O(d^2n^2)+O(d^3n)$ & $O(d^3n)$ & 0         & 0      \\ \hline
		\texttt{SecMatEigen} & $O(d^2n)$ & 0         & $O(d^2n^2)+O(d^3n)$ & $O(d^3n)$ & 0         & 0      \\ \hline
	\end{tabular}
\end{table*}

 Few works in SMC pay their attention to computing nonlinear computation losslessly. In this case, we here only give the comparison on $\texttt{SecCmp}$ and $\texttt{SecDiv}$ as shown in Table \ref{tab:CompOnCmp}, where $n$ is set to be $2$ for the fair comparison. Compared to previous works, our scheme has lower interaction rounds.

\begin{table}[]
	\caption{Comparison of communication complexities. \\(Here, $l$ and $k$ is the bit-width and byte size of \textit{share})}
	\centering
	\label{tab:CompOnCmp}
	\begin{tabular}{|l|c|c|c|c|}
		\hline
		\multirow{2}{*}{Scheme}                       & \multicolumn{2}{c|}{\texttt{SecCmp}} & \multicolumn{2}{c|}{\texttt{SecDiv}} \\ \cline{2-5} 
		& Rounds & \begin{tabular}[c]{@{}c@{}}Comm\\ (bits)\end{tabular} & Rounds & \begin{tabular}[c]{@{}c@{}}Comm\\ (bits)\end{tabular} \\ \hline
		
		\cite{nishide2007multiparty}          & $15$         & $279l+5$         & -          & -              \\ \hline
		\cite{huang2019lightweight}           & $l+3$        & $10l-2$          & -          & -              \\ \hline
		\cite{wagh2020falcon}				  & $log_{2}l+2$ & $2l$				& $7+(5+log_{2}l)l$ & $4kl+7k$ \\ \hline
		\cite{xiong2020efficient}             & $3$          & $2l+2$           & $3$          & $6l$             \\ \hline
		Ours                                  & $2$          & $6l$             & $2$          & $10l$            \\ \hline
	\end{tabular}
\end{table}

\section{A case Study: convolutional neural network}\label{sec:CaseStudy}

\subsection{Background}

 The rapid development of Convolutional Neural Network (CNN) is one of the most important research achievements in recent years \cite{qiu2020review,zeng2020data}. Typically, the feature extraction can be replaced by a series of simple operations in CNN models. As an example of Deep Learning as a Service (DLaaS), it is a promising application paradigm for resource-limited users to extract CNN features by utilizing the AI models trained and deployed on the cloud servers. However, the input data of CNN model, e.g., human face, is often highly sensitive. Therefore, supporting the computation of the CNN model with the encrypted inputs attracts more and more researchers. Similar to previous works, we focus on the security of input, rather than the model itself \cite{ qiu2020adversarial, li2020intelligent}.

 Early works tried to use HE tools to cope with this fundamental problem. However, due to plenty of nonlinear and comparison operations, the efficiency of these works is unacceptable \cite{gilad2016cryptonets}. Two main kinds of strategies are proposed to alleviate this problem. One is to seek approximate algorithms to replace complex operations \cite{mohassel2017secureml}, and the other tries to use the technology in SMC, such as GC, to assist in the computations \cite{juvekar2018gazelle}. Notably, such schemes are time-consuming and easy to cause precision loss. In this case, the schemes completely based on secret sharing become the most promising solution at present. Typically, compared to the above kinds of schemes, the work in \cite{huang2019lightweight} is hundreds of times more efficient. Yet, as described in the above sections, these schemes still fail to cope with the nonlinear operations, which inevitably causes the loss of precision and time.

\subsection{Computations in CNN}

 Convolutional Neural Network (CNN) includes a sequence of layers that compute and transform the input into the output. In general, a CNN model is composed of the Convolutional layer (Conv), Activation Layer (AL), Pooling layer (Pool), and Fully Connected layer (FC). Similar to previous works, we here only describe the inference process of CNN with known parameters (i.e., weights and biases).

 For FC and Conv layers, the inference process only involves linear computations since the parameters are known. Thus, when the inputs are split to additive \textit{shares} and stored in two servers, the secure inference process is just the same as that in plaintext state. For AL, there are two kinds of mainstream activation functions: the first one involves comparison, e.g., $ReLU$ function, and the other involves non-linear functions and division operation, e.g., $sigmoid$ function. All of the operations can be losslessly supported by our protocol. With parallelization, the interaction rounds are irrelevant to the size of the input. The Pooling layer tries to compress the numbers in the fixed-size block to a number. Mean-pooling and max-pooling are two widely used schemes. As the size of the block is a hyper-parameter known to both servers, the mean-pooling only involves linear calculations. For max-pooling, the servers try to find the maximum number in the block. To get the maximum value, the servers can compare each pair of numbers in the block. Since the block utilized is usually small, the actual time of comparisons is low. For instance, only $6$ times of comparison is needed in a $2\times{}2$ block. Due to parallelization, the interaction rounds required are also irrelevant to the input and block size.

\begin{figure*}[ht]
	\centering
	\includegraphics[width=1.0\linewidth]{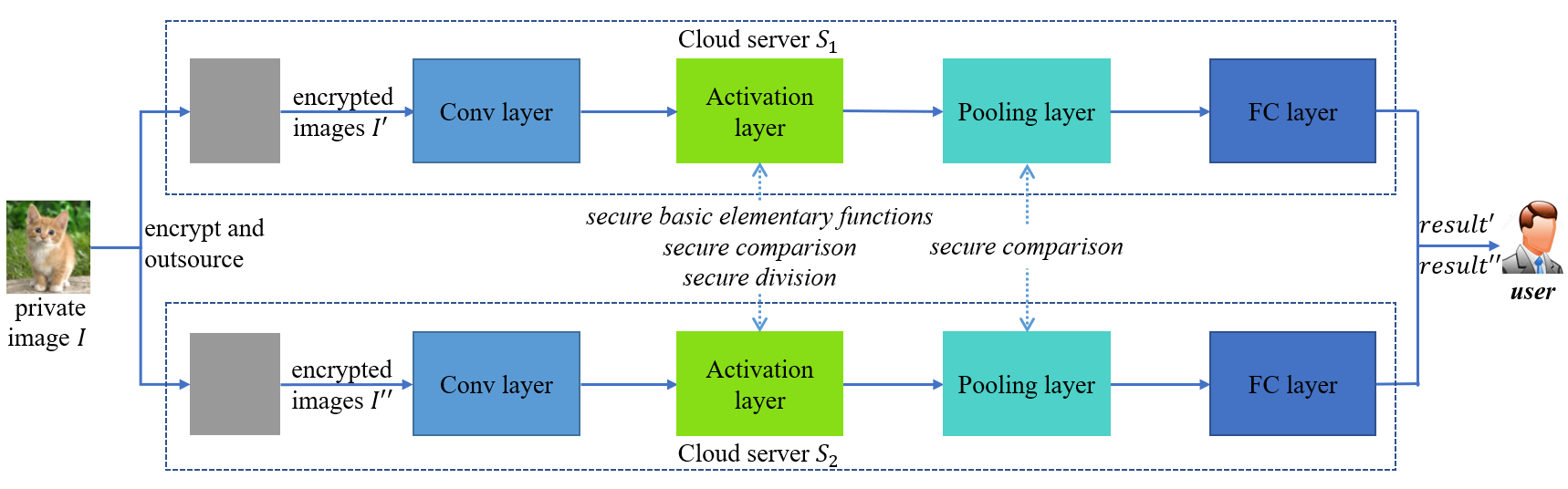}
	\caption{Secure inference process of convolution neural network}
	\label{fig:CNNInfer}
\end{figure*}

 To sum up, all the common operations of the CNN model can be supported by our protocols directly. A framework of such applications is illustrated in Fig. \ref{fig:CNNInfer}. To show the advantage in practice, several common models are utilized to evaluate the efficiency.

\subsection{Efficiency evaluation}

 To be consistent with previous works \cite{huang2019lightweight}, the $(2,2)$-threshold model is tested. We evaluate our protocols with three CNN models, including Network \RNum{1}: $1$-Conv and $2$-FC with square activation from \cite{gilad2016cryptonets}, Network \RNum{2}: $2$-Conv and $2$-FC with ReLU and max-pooling from \cite{liu2017oblivious}, and Network \RNum{3}: $1$-FC and $7$-Conv with ReLU from \cite{liu2017oblivious}. All the experiments in the owner side (i.e., trustworthy party) are executed in a machine with Intel Core i5-8250u CPU @ 1.6GHZ and 16 GB memory. The experiments in the cloud side are executed on two servers in the LAN setting. Each server is equipped with Intel Core i7-9700 CPU @ 3GHZ and 16 GB memory. The experiment results are shown in Table \ref{tab:InferCons1}, \ref{tab:InferCons2}, and \ref{tab:InferCons3}.

\begin{table}[]
	\centering
	\caption{Inference consumption of CNN model \RNum{1}}
	\label{tab:InferCons1}
	\begin{tabular}{|l|c|c|c|c|}
		\hline
		\multirow{2}{*}{} & \multicolumn{2}{c|}{Runtime(s)} & \multicolumn{2}{c|}{Message sizes(MB)} \\ \cline{2-5} 
		& Offline & Online & Offline & Online \\ \hline
		CryptoNets \cite{gilad2016cryptonets} 	& 0 	& 297.5 	& 0 	& 372.2 \\ \hline
		MiniONN \cite{liu2017oblivious} 			& 0.88  & 0.4     	& 3.6   & 44   \\ \hline
		Gazelle \cite{juvekar2018gazelle} 		& 0 	& 0.03 	  	& 0     & 0.5    \\ \hline
		Huang \cite{huang2019lightweight}   		& 0.0009& 0.002   	& 0.022 & 0.015   \\ \hline
		Ours    							& 0.00005   & 0.0024   & 0.129    & 0.015   \\ \hline
	\end{tabular}
\end{table}

\begin{table}[]
	\centering
	\caption{Inference consumption of CNN model \RNum{2}}
	\label{tab:InferCons2}
	\begin{tabular}{|l|c|c|c|c|}
		\hline
		\multirow{2}{*}{} & \multicolumn{2}{c|}{Runtime(s)} & \multicolumn{2}{c|}{Message sizes(MB)} \\ \cline{2-5} 
		& Offline & Online & Offline & Online \\ \hline
		MiniONN \cite{liu2017oblivious}  & 3.58    & 5.74     & 20.9    & 636.6   \\ \hline
		Gazelle \cite{juvekar2018gazelle}& 0.481    & 0.33   & 47.5     & 22.5    \\ \hline
		Huang \cite{huang2019lightweight}& 0.09    & 0.21   & 1.57    & 0.99   \\ \hline
		Ours    						& 0.002   & 0.012   & 4.71    & 0.59   \\ \hline
	\end{tabular}
\end{table}

\begin{table}[]
	\centering
	\caption{Inference consumption of CNN model \RNum{3}}
	\label{tab:InferCons3}
	\begin{tabular}{|l|c|c|c|c|}
		\hline
		\multirow{2}{*}{} & \multicolumn{2}{c|}{Runtime(s)} & \multicolumn{2}{c|}{Message sizes(MB)} \\ \cline{2-5} 
		& Offline & Online & Offline & Online \\ \hline
		MiniONN \cite{liu2017oblivious} 	 & 472     & 72     & 3046    & 6226   \\ \hline
		Chameleon \cite{riazi2018chameleon} & 22.97    & 29.7   & 1210     & 1440    \\ \hline
		Huang \cite{huang2019lightweight} & 0.62    & 1.55   & 10.57    & 6.61   \\ \hline
		Ours    						 & 0.0142   & 0.019   & 31.69    & 2.65   \\ \hline
	\end{tabular}
\end{table}

 Except for the previous work \cite{huang2019lightweight} that is also based on additive secret sharing, all other works exist obvious efficiency deficiencies. Moreover, since the comparison protocol in \cite{huang2019lightweight} is based on bit-decomposition technology \cite{damgaard2006unconditionally}, the efficiency in Network \RNum{2} and \RNum{3} is also inferior to our protocols. Here, since we focus on the computation on decimals (e.g., 24 bytes per number), the \emph{offline} communication size consumption is higher. The difference between \emph{offline} time consumption is more due to the equipment we use. However, the advantages in \emph{online phase} can demonstrate the superiority of our protocols.

 Since the proposed protocols include a whole set of basic computations, our work can better fit the various neural network models. Unlike the works in federal learning, we use additive secret sharing to protect the privacy input and even the parameters in models. Thus, our scheme can better resist attacks in federal learning (e.g., model inversion attack). We believe the proposed protocols are useful for the deep-learning platforms which care about privacy. The application of our protocol to the training process and more complex models would be an interesting follow-up.

\section{Conclusion and future work}\label{sec:conclu}

 In this paper, we focus on the outsourced computation problem. With the help of secret sharing technology, we construct a series of basic computation protocols on numbers and matrices. We do not limit the number (e.g., $n$) of participating servers and ensure the security of original data even if $n-1$ servers collude. The security of protocols is infinitely close to FHE schemes; however, the efficiency, precision, and scalability are far better. We believe that the protocols provide a potential tool for plenty of secure outsourced tasks in the cloud computing scene. Due to the generality of secret sharing technology, traditional SMC tasks will also benefit from these protocols. 

 In the future, it is meaningful to construct various secure outsourcing schemes using the proposed protocol and to design more protocols on matrices. Moreover, many challenging problems can be further considered. For instance, is it feasible to let servers undertake the tasks during the \emph{offline phase} securely and efficiently?

\section*{Acknowledgements}
 This work is supported in part by the Jiangsu Basic Research Programs-Natural Science Foundation under grant numbers BK20181407, in part by the National Natural Science Foundation of China under grant numbers U1936118 and 61672294, in part by Six Peak Talent project of Jiangsu Province (R2016L13), Qinglan Project of Jiangsu Province, and ``333'' project of Jiangsu Province, in part by the Priority Academic Program Development of Jiangsu Higher Education Institutions (PAPD) fund, in part by the Collaborative Innovation Center of Atmospheric Environment and Equipment Technology (CICAEET) fund, China, in part by Guangxi Key Laboratory of Trusted Software under the grant number KX202044. Zhihua Xia is supported by BK21+ program from the Ministry of Education of Korea.

\bibliographystyle{IEEEtran}

\bibliography{HowToCalculateOnSecrets}

\begin{IEEEbiography}[{\includegraphics[width=1in,height=1.25in,clip,keepaspectratio]{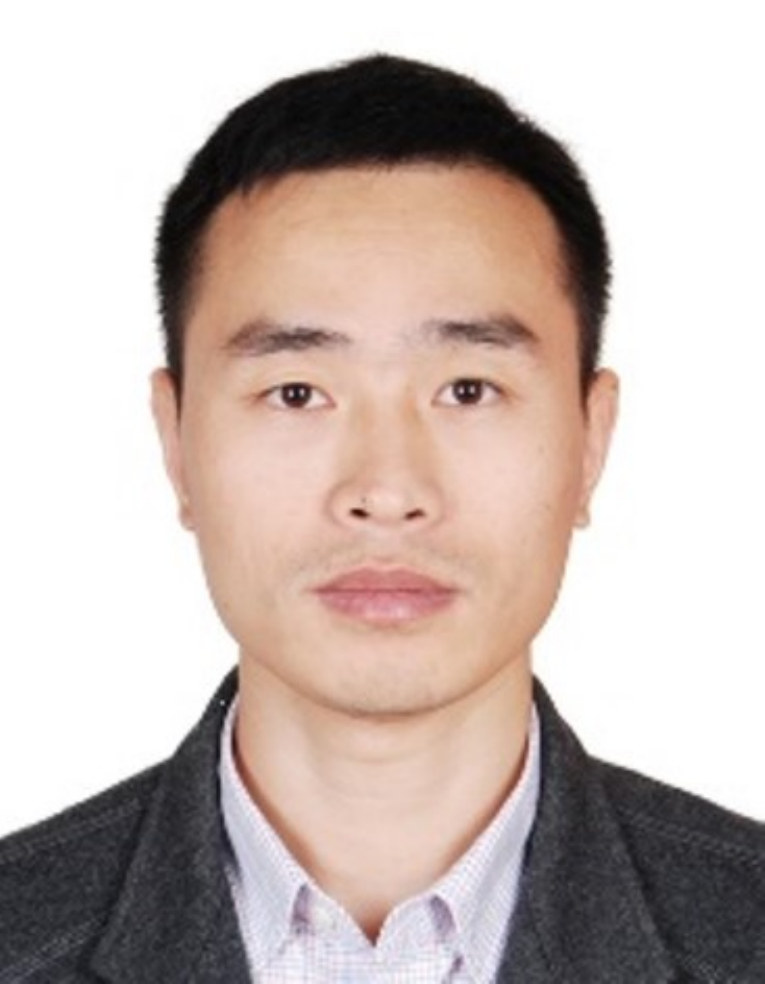}}]{Zhihua Xia }
		received his Ph.D. degree in computer science and technology from Hunan University, China, in 2011. He is currently a professor with the College of Cyber Security, Jinan University, China. He was a visiting scholar at New Jersey Institute of Technology, USA, in 2015, and was a visiting professor at Sungkyunkwan University, Korea, in 2016. He serves as a managing editor for International Journal of Autonomous and Adaptive Communications Systems. His research interests include cloud computing security and digital forensic. He is a member of the IEEE.
\end{IEEEbiography}

\begin{IEEEbiography}[{\includegraphics[width=1in,height=1.25in,clip, keepaspectratio]{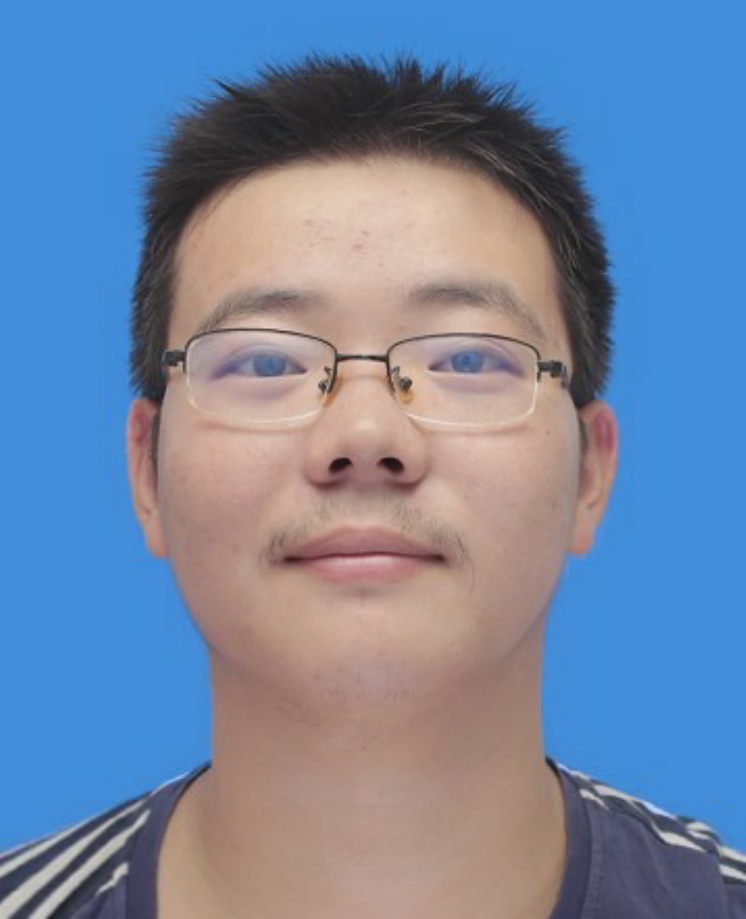}}]{Qi Gu}
	is currently pursing his master degree in the School of Computer and Software, Nanjing University of Information Science and Technology, China. His research interests include functional encryption, image retrieval and nearest neighbor search.
\end{IEEEbiography}

\begin{IEEEbiography}[{\includegraphics[width=1in,height=1.25in,clip, keepaspectratio]{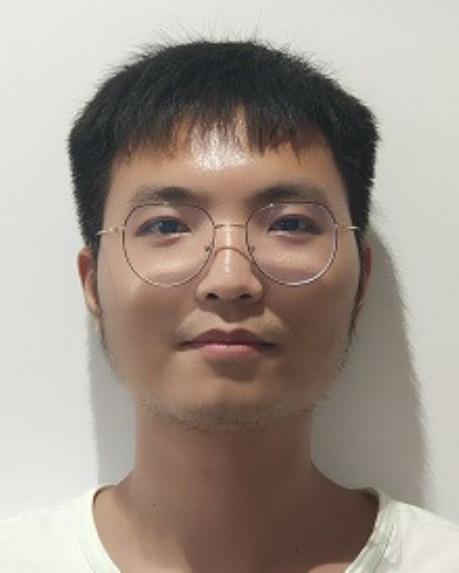}}]{Wenhao Zhou}
	is currently pursing his master degree in the School of Computer and Software, Nanjing University of Information Science and Technology, China. His research interests include secure multiparty computation and privacy-preserving computation.
\end{IEEEbiography}

\begin{IEEEbiography}[{\includegraphics[width=1in,height=1.25in,clip,keepaspectratio]{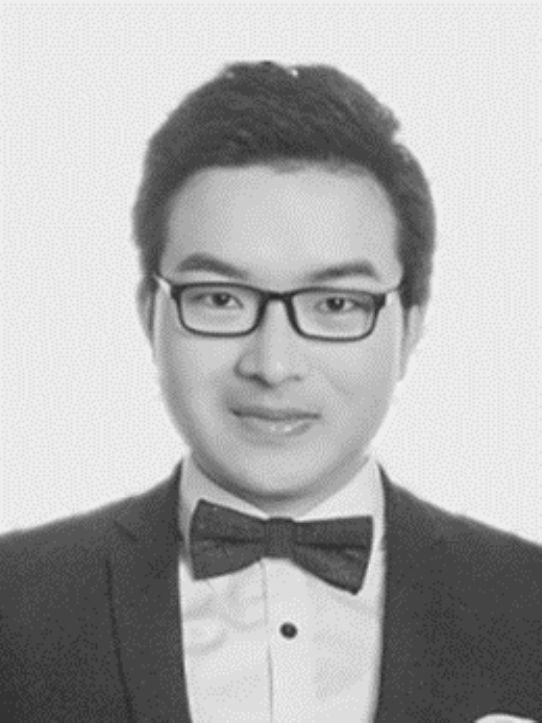}}]{Lizhi Xiong}
	received Ph.D. degree in Communication and Information System from Wuhan University, China in 2016. From 2014 to 2015, he was a Joint-Ph.D. student with Electrical and Computer Engineering, New Jersey University of Technology, New Jersey, USA. He is currently an Associate Professor with School of Computer and Software, Nanjing University of Information Science and Technology, Nanjing, China. His main research interests include privacy-preserving computation, information hiding, and multimedia security.
\end{IEEEbiography}

\begin{IEEEbiography}[{\includegraphics[width=1in,height=1.25in,clip,keepaspectratio]{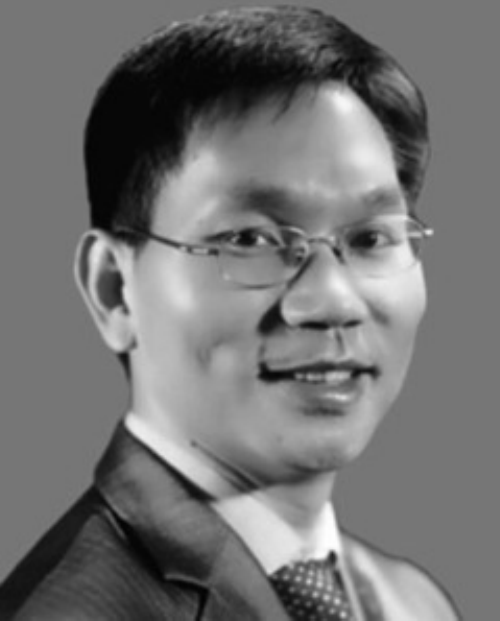}}]{Jian Weng}
	received the Ph.D. degree in computer science and engineering from Shanghai Jiao Tong University, Shanghai, China, in 2008. He is currently a Professor and the Dean with the College of Information Science and Technology, Jinan University, Guangzhou, China. His research interests include public key cryptography, cloud security, and blockchain. He was the PC Co-Chairs or PC Member for more than 30 international conferences. He also serves as an Associate Editor for the IEEE TRANSACTIONS ON VEHICULART ECHNOLOGY. 
\end{IEEEbiography}

\begin{IEEEbiography}[{\includegraphics[width=1in,height=1.25in,clip,keepaspectratio]{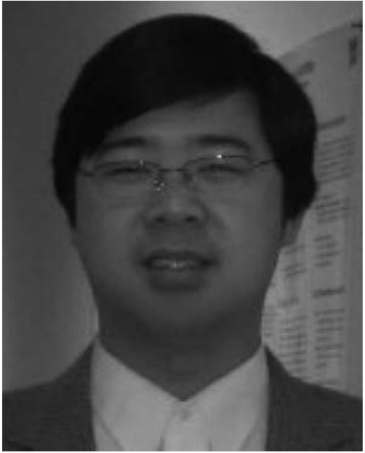}}]{Neal N. Xiong}
	received the first Ph.D. degree in sensor system engineering from Wuhan University, Wuhan, China, and the second Ph.D. degree in dependable sensor networks from the Japan Advanced Institute of Science and Technology, Nomi, Japan. He is currently a Professor with the College of Intelligence and Computing, Tianjin University, Tianjin, China. He was with Georgia State University, Atlanta, GA, USA; Wentworth Technology Institute, Boston, MA, USA; and Colorado Technical University, Colorado Springs, CO, USA (a Full Professor for about five years) for about ten years. He was with Northeastern State University, Tahlequah, OK, USA. He has published over 200 international journal papers and over 100 international conference papers. His current research interests include cloud computing, security and dependability, parallel and distributed computing, networks, and optimization theory.
\end{IEEEbiography}

\newpage
\mbox{}
\newpage

\section*{Appendix}

\subsection{Power of rational exponent}

Since the pow computation with irrational number exponent is still not clear, we just consider the exponent $a$ in $\mathbb{Q}$, which is also the superset of the finite field $\mathbb{F}$. In this case, the power function will become a multi-valued function. In the real task, we believe that only the principal value is the desired result. Thus, when the base number is positive, the corresponding principal value will be gotten. However, when the base number is negative, the \emph{complex number} will be inevitably involved. Whatever, it is a problem that should be determined in the specific task. Considering the identity

\begin{equation}
\label{eq:SecPow-Q}
\begin{array}{c}
u^{a} = (-1)^{a}|u|^{a},
\end{array}
\end{equation}

\noindent where $u$ is negative, the choice of value is determined by the choice of $(-1)^{a}$ and its value can be defined in the specific task (e.g., define $(-1)^{\frac{1}{3}}=-1$). The detailed protocol \texttt{SecPRE} is shown in Algorithm \ref{alg: SecPow2}.

\begin{algorithm}[h]
	\caption{Secure Power with Rational Exponent protocol $\texttt{SecPRE}$}
	\label{alg: SecPow2}
	\begin{algorithmic}[1]
		\REQUIRE
		$S_i$ has $\ass{x}_i$ and a public exponent $a$ with $x\neq{}0, a\in \mathbb{Q}$.
		\ENSURE
		$S_i$ gets $\ass{x^a}_i$.
		
		\textbf{Offline Phase} :
		
		\STATE $\mathcal{T}$ generates enough random numbers that the sub-protocols use and sends them to $S_i$.
		
		\textbf{Online Phase} :
		
		\STATE $S_i$ collaboratively compute $\texttt{SecCmp}(\ass{x}_i, 0)$ to judge the sign of $x$.
		
		\STATE $S_i$ collaboratively compute $\mss{x}_i = \texttt{SecAddResh}(\ass{x}_i)$.
		
		\STATE $S_i$ computes $\mss{|x|^a}_{i} = |\mss{x}_{i}|^a$.
		
		\STATE $S_i$ collaboratively compute $\ass{|x|^a}_i = \texttt{SecMulResh}($ $\mss{|x|^a}_i)$.
		
		\STATE $S_i$ computes $\ass{x^a}_i = (-1)^a \times \ass{|x|^a}_i$ if $x$ is negative.
		
	\end{algorithmic}
\end{algorithm}

\subsection{Inverse Trigonometric Functions}

The inverse trigonometric functions are $arcsin$, $arccos$, $arctan$, $arccot$, $arcsec$, and $arccsc$. As each one could compute from $arctan$ under the transformation of input, we focus on the secure computation protocol on $arctan$: $\ass{arctan(x)}_i \leftarrow \ass{x}_i$.

To our knowledge, no equation could directly transform input to the output in ASS or MSS format. Hence, we construct the output that satisfies our demand directly. To this end, let us consider the equation

\begin{equation}
\label{eq:Arctan}
\begin{array}{c}
arctan(u) + arctan(\frac{x-u}{1+ux}) = arctan(x),
\end{array}
\end{equation}

\noindent where $u\times{\frac{x-u}{1+ux}}$ should be less than $1$. The $u\times{\frac{x-u}{1+ux}}$ will not be equal to $1$ for any $u$ and $x$ in $\mathbb{R}$. If it is greater than $1$, then $\pm\pi$ should be added based on the sign of $u$ (or $\frac{x-u}{1+xu}$).

First, we consider $arctan$ in the $(2,2)$-threshold model: two servers own $\ass{x}_1$ and $\ass{x}_2$. We wish the server can own $u$ and $\frac{x-u}{1+ux}$ independently so that the output can be locally computed and stored in ASS format. The problem becomes how to transform the additive \textit{share} of $x$ to $u$ and $\frac{x-u}{1+ux}$ in security.

To achieve the above goal, the following process can be performed: $S_1$ generates a random number $u$; $S_i$ collaboratively compute the $\ass{x-u}_i$ and $\ass{1+ux}_i$; then, the division is secretly computed; finally, $S_1$ sends $\ass{\frac{x-u}{1+ux}}_1$ to $S_2$, then the transformation is completed. Since each server still does not know anything about $x$, the above process can be regarded as a $(2,2)$-threshold scheme. In this case, we should expand the process to the $(n,n)$-threshold situation.

It is difficult to expand the Equation (\ref{eq:Arctan}) directly. Therefore, we use an alternative method that transforms the computation in $(n,n)$-threshold to $(n-1,n-1)$-threshold process. Specifically, $n$ servers can be divided into two parts. Through the above process, $n-1$ servers get one \textit{share} of $arctan(x)$, and the remaining one gets another \textit{share}. Accordingly, we can further decrease $n-1$ to $n-2$ until it becomes $(2,2)$-threshold. The above process can be designed as shown in algorithm \ref{alg: PartArcTan} and \ref{alg: ArcTan}.

\begin{algorithm}[t]
	\caption{One iteration of Secure $arctan$ protocol}
	\label{alg: PartArcTan}
	\begin{algorithmic}[1]
		\REQUIRE
		$S_i$ has $\ass{x}_i$, $S_j$ has $\ass{u}_j$. $(i\in[l,n], j\in[l+1,n])$
		\ENSURE
		$S_l$ gets $arctan(\frac{x-u}{1+xu})$.
		
		\textbf{Offline Phase} :
		
		\STATE $\mathcal{T}$ generates enough random numbers that the sub-protocols use and sends them to $S_i$.
		
		\textbf{Online Phase} :
		
		\STATE $S_i$ computes $\ass{x-u}_i = \ass{x}_i - \ass{u}_i$, here $\ass{u}_l = 0$.
		
		\STATE $S_i$ collaboratively compute $\ass{xu}_i = \texttt{SecMul}(\ass{x}_i, \ass{u}_i)$.
		
		\STATE $S_i$ collaboratively compute $\ass{\frac{x-u}{1+xu}}_i = \texttt{SecDiv}(\ass{x-u}_i, \ass{1+xu}_i)$.
		
		\STATE $S_j$ sends $\ass{\frac{x-u}{1+xu}}_j$ to $S_l$.
		
		\STATE $S_i$ collaboratively compute $\texttt{SecCmp}$$(\texttt{SecMul}$$(\ass{\frac{x-u}{1+xu}}_i$, $\ass{u}_i), 1)$, here $\ass{\frac{x-u}{1+xu}}_j=0$.
		
		\STATE If $u\times{}\frac{x-u}{1+xu}$ more than $1$, then $S_l$ computes $arctan(\frac{x-u}{1+xu})$ $=$ $arctan(\frac{x-u}{1+xu}) - sgn(\frac{x-u}{1+xu})\pi$.
		
	\end{algorithmic}
\end{algorithm}

\begin{algorithm}[t]
	\caption{Secure $arctan$ protocol $\texttt{SecArctan}$}
	\label{alg: ArcTan}
	\begin{algorithmic}[1]
		\REQUIRE
		$S_i$ has $\ass{x}_i$.
		\ENSURE
		$S_i$ gets $\ass{arctan(x)}_i$.
		
		\textbf{Offline Phase} :
		
		\STATE $\mathcal{T}$ generates enough random numbers that the sub-protocols use and sends them to $S_i$.
		
		\textbf{Online Phase} :
		
		\STATE $S_i$ sets $\ass{u^1}_i = \ass{x}_i$.
		
		\FOR{$l = 2:n$}
		
		\STATE $S_j$ generates random numbers $\ass{u^l}_j$, $j\in[l,n]$.
		
		\ENDFOR
		
		\FOR{$l = 1:n-1$}
		
		\STATE $S_j$ collaboratively computes Algorithm \ref{alg: PartArcTan} with input $\ass{u^l}_j$ and $\ass{u^{l+1}}_j$, $j\in[l,n]$.
		
		\STATE $S_l$ gets the corresponding $\ass{arctan(x)}_l$.
		
		\ENDFOR
		
		\STATE $S_n$ computes $\ass{arctan(x)}_n$ $=$ $arctan(u^n)$.
		
	\end{algorithmic}
\end{algorithm}

Notably, Algorithm \ref{alg: PartArcTan} can be executed in parallel in that the inputs are pre-generated in $\texttt{SecArctan}$. Thus, the round of interaction is the same as that in (2,2)-threshold. In theory, the rounds of communication is $5$ and the number of involved additive \textit{share} is $\frac{17}{3}n^{3}+\frac{1}{2}n^{2}-\frac{37}{6}n+16$.

\end{document}